# A semi-empirical integrated microring cavity approach for 2D material optical index identification at 1.55 μm

RISHI MAITI,1 ROHIT A. HEMNANI,1 RUBAB AMIN,1 ZHIZHEN MA,1 MOHAMMAD H. TAHERSIMA,1 TOM A. EMPANTE,2 HAMED DALIR,3 RITESH AGARWAL,4 LUDWIG BARTELS,2 AND VOLKER J. SORGER1,*

1Department of Electrical and Computer Engineering, George Washington University, Washington, DC 20052, USA
2Chemistry and Materials Science and Engineering, University of California, Riverside, California 92521, USA
3Omega Optics, Inc. 8500 Shoal Creek Blvd., Bldg. 4, Suite 200, Austin, TX 78757, USA
4Department of Materials Science and Engineering, University of Pennsylvania, Philadelphia, PA 19104, USA
*Corresponding author: sorger@gwu.edu

**Atomically thin two-dimensional (2D) materials provide a wide range of basic building blocks with unique properties, making them ideal for heterogeneous integration with a mature chip platform for advances in optical communication technology. Control and understanding of the precise value of the optical index of these materials, however, is challenging, due to the small lateral flake dimension. Here we demonstrate a semi-empirical method to determine the index of a 2D material ($n_{MoTe2}$ of 4.36+0.011$i$) near telecommunication-relevant wavelength by integrating few layers of MoTe$_2$ onto a micro-ring resonator. The placement, control, and optical-property understanding of 2D materials with integrated photonics paves a way for further studies of active 2D material-based optoelectronics and circuits.**

The ever-increased demand for high performance computing requires efficient on-chip interconnects [1]. Hence, intensive efforts have been devoted to achieving compact photonic components such as light source, detector, and modulator which is challenging to realize within one single monolithic platform. In contrast, heterogeneous integration of two-dimensional (2D) material on Silicon (Si) offers a promising solution since it provides a highly tunable and functional platform for a variety of optoelectronic devices [2-8]. To proper aid the design and anticipate device performance, knowledge of the complex refractive index is fundamentally important.  Recently, few groups have measured the optical index of transition metal dichalcogenides (TMDCs) using ellipsometry [9,10]. However, there are two challenges with using ellipsometry to determine the refractive index of TMDCs and devices fabricated thereof; a) typical optoelectronic device dimensions being on the order of micrometers are too small to be probed by ellipsometry, and b) the state of the art of TMDCs synthesis does not allow to growth sufficiently uniform large films with high quality yet. Even if that could be solved, the thermal budget may not sufficient to allow for direct CVD growth on the actual device or circuit substrate. Also, even if wafer-scale growth would be possible, the subsequent required etch step will impact the surface-sensitive optical index of these 2D materials. As a result, determining the refractive index of the TMDCs in-situ is a requirement which we address here using integrated photonic microcavities.

2D materials are actively investigated due to their interesting properties in various fields of physics, chemistry, and materials science, starting with the isolation of graphene from graphite [11]. While graphene shows many exceptional properties, its lack of an electronic bandgap has stimulated the search for alternative 2D materials with semiconducting characteristics [12-16]. TMDCs, which are semiconductors of the type MX$_2$, where M is a transition metal atom (such as Mo or W) and X is a chalcogen atom (such as S, Se or Te), provide a promising alternative. 2D TMDCs exhibit unique physical properties such as indirect to direct bandgap transition [17], quantum confinement with photonic crystal cavity [18-21], and high exciton binding energies [22] as compared to their bulk counterparts, making them versatile building blocks for optical modulators or reconfiguration in a Si-based hybrid platform. Here, we demonstrate heterogeneous integration of few layers of MoTe$_2$ on Si MRR platform. The interaction between the TMDC and the Si MRR provides novel tunable coupling phenomena that can be tuned from over coupling to under coupling regime via critical coupling condition by means of altering the rings effective index via the integration of TMDCs. We analyze the coupling physics and extract fundamental parameters such as quality factor, visibility, transmission at resonance etc., as a function of MoTe$_2$ coverage on the ring. We find a critical-coupling coverage-ratio value (~10%) for a given ring resonator which is relevant for device functionality. Furthermore, we determine the index ($n_{MoTe2}$ = 4.36+0.011$i$) of the few layered MoTe$_2$ at 1.55 μm wavelength in a semi-empirical approach using the ring resonator as an index sensor platform.

In order to obtain an understanding of phase modulation in heterogeneous integrated systems, it is important to understand the interaction between TMDCs and a MRR. Unlike other larger-bandgap TMDCs (for example, MoS$_2$, WS$_2$ MoSe$_2$, WSe$_2$), the optical bandgap of few layered MoTe$_2$ (~0.9 eV) is smaller than the bandgap of Si (~1.1 eV), allows a facile integration with Si photonic platform. Here, the aim is to improve the waveguide bus-to-ring coupling efficiency by shifting the phase through introducing a few-layered of MoTe$_2$ flakes on top of the resonator (Fig. 1a, b & c). By varying the MoTe$_2$ coverage length, we observe a tunable coupling response. To achieve TMDCs-loading of the rings, we utilize a precise and cross-contamination free transfer using the 2D printer method recently developed [23]. The transmission spectra before and after MoTe$_2$ transfer reveal definitive improvement of coupling in terms of visibility defined as the transmission amplitude ratio ($T_{max}/T_{min}$) upon TMDCs loading (Fig. 1d). This hybrid device shifts towards the critical coupling regime as compared to before transfer where it was over coupled.

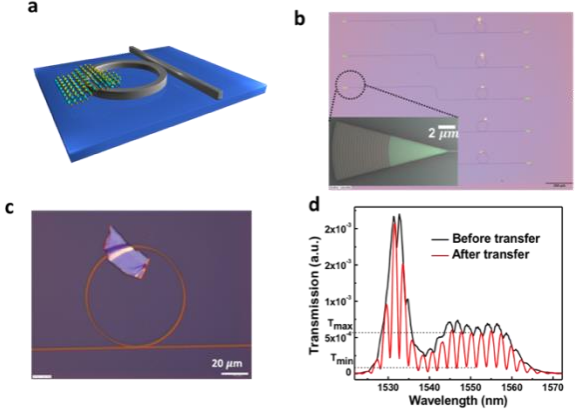

**Fig. 1.** TMDCs loaded micro-ring resonator (MRR) (a) schematic and (b) optical microscope image of a tapedout Si photonics chip with grating coupler. Optical micrograph of a grating coupler designed for transverse magnetic (TM) mode at 1550 nm in the field of a view of 100x objective (inset). (c) Zoomed image of a MRR (R= 40 μm & W= 500 nm) covered by a MoTe$_2$ flake of length (*l*) precisely transferred using our developed 2D printer technique. (d) Transmission output before and after the transfer of MoTe$_2$ showing improvement of coupling efficiency as it brings the device close to critically coupled regime after the transfer of the TMDCs layer.

We fabricated a set of ring resonators with different percentages of MoTe$_2$ coverage between 0% and 30% (Fig. 2) which allowed us to extract different parameters to understand coupling physics. The cavity quality (*Q*) factor is found to decrease from 1600 to 900 as the ring coverage is raised from 0 to 27% (Fig. 2a). We attribute this as gradual increase of loss arising from both MoTe$_2$ absorption near its band edge corresponding to indirect bandgap of 0.88 eV [24], and the small impedance mismatch between bare and TMDC-covered sections of the rings. In contrast to the monotonic behavior of the quality factor, the minimum transmission ($T_{min}$) at resonant wavelength initially decreases until 10% coverage is reached and then increases for higher coverages (Fig. 2b). The visibility ($T_{max}/T_{min}$) shows the opposite trend being maximal near 10% of TMDC coverage (Fig. 2c). In combination, these findings indicate tunability of the coupling condition by means of varying TMDC coverage.

The performance of a ring resonator is determined by two coefficients: the self-coupling coefficient (*r*), which specifies the fraction of the light transmitted on each pass through the coupler; and the round-trip transmission coefficient (*a*), which specifies the fraction of the light transmitted per pass around the ring. For the critical coupling condition, i.e. when the coupled power is equal to the power loss in the ring $1-a^2=k^2$ or $a=r$, (*k*=cross coupling coefficient), the transmission at resonance becomes zero. At this point, the difference (|*a-r*|) is found to be minimum at ~10% of coverage (Fig. 2d) suggesting close to critical coverage since |*a-r*| is inversely proportional to the square root of visibility term [25]. Thus, the coupling condition is tunable from the over coupled regime (*r*<*a*) to under coupled (*r*>*a*) via the critically coupled regime (*r*=*a*) as a function of TMDCs coverage. Being able to operate at critical coupling is important for active device functionality; for instance, the extinction ratio of an MRR-based electro optic modulator is maximized at critical coupling [26,27].

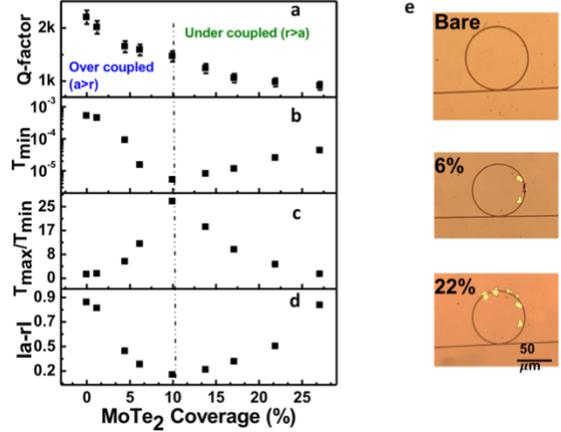

**Fig. 2.** TMDCs loaded tunable coupling effect. Variation of (a) *Q*-factor; (b) Minimum transmission, $T_{min}$; (c) Visibility ($T_{max}/T_{min}$); (d) Difference between the self-coupling coefficient and round-trip transmission coefficient, |*a-r*|; as a function of MoTe$_2$ coverage. The monotonic decrease of *Q*-factor suggests increasing loss for higher coverage, i.e. more transferred TMDCs. Tunability of coupling effect i.e. transition from over-coupled to under-coupled regime via critically-coupled condition (dashed vertical line) is evident from the variation of $T_{min}$, $T_{max}/T_{min}$ & |*a-r*|. (e) The corresponding figures for different coverages (bare ring, 6% & 22%) showing the advantages of precise transfer by 2D transfer methods.

In order to understand, the coupling mechanism of a ring resonator, it is important to extract and distinguish coupling coefficients (*a* and *r*), as they are governed by different factors in design and fabrication. However, it is not possible to decouple both the coefficients without additional information, since *a* and *r* can be interchanged (eq.1). The transmission from an all-pass MRR (Fig. 3a) is given by,

$$T_n = \frac{a^2+r^2-2ar\cos\varphi}{1+r^2a^2-2ar\cos\varphi} \quad (1)$$

where, $\varphi$ is the round-trip phase shift, *a* is round-trip transmission coefficient related to the power attenuation coefficients by,

$$a^2 = \exp(-\alpha_{Si}(2\pi R - l)) \times \exp(-\alpha_{TMD-Si} \cdot l) \quad (2)$$

where *l* = TMDC coverage length, $\alpha_{Si}$ and $\alpha_{TMDC-Si}$ are the linear propagation losses for Si waveguide and TMDC-transferred portion of the Si waveguide in the ring, respectively. We find the propagation loss for Si and TMDC-Si to be 0.008 dB/μm and 0.4 dB/μm (Fig. 3b), respectively via the cutback method at 1550 nm.

Inserting these values into (2), we find the round-trip transmission coefficients (*a*) to be tuned as a function of TMDCs coverage (Fig. 3c). The variation of *a* from 0.97 to 0.01 as a function of coverage confirms the transition from over-coupled to under-coupled regime since *a* = 1 suggests that there is no loss in the ring. Hence, the loss tunability in MRRs can be manipulated accordingly by controlling the coverage length [28]. The MRR transmission at resonance leads to the form, $T_{n,res} = \left(\frac{a-r}{1-ar}\right)^2$, therefore, the critical-coverage ($a=r$) anticipates zero-output transmittance. Hence, for a given MRR (fixed $k_2$), the critical coverage value could be determined provided lossless coupling ($r_2+k_2=1$) (Fig. 3d). Si-based MRRs provide a compact and ultra-sensitive platform as refractive index sensor to find an unknown index of the 2D materials at telecom wavelength (1.55 μm). In essence, the shift in resonant wavelength can be used to sense the optical properties effecting entities on the silicon core or the cladding [29]. We observe a gradual resonance red-shift from bare to increased coverage of MoTe₂ of 4.5%, 10% and 17%, respectively (Fig. 4a).

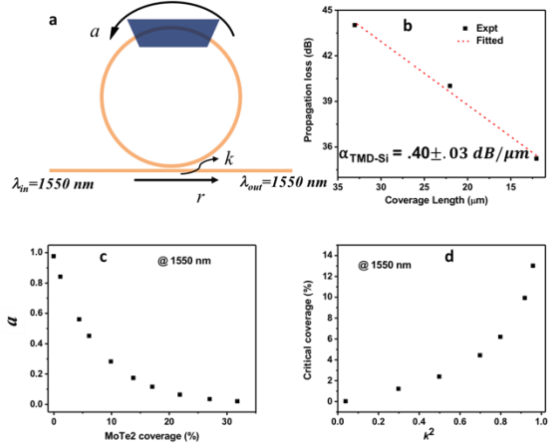

**Fig. 3.** Coupling coefficients for TMDCs integrated hybrid Si MRR a) Schematic representation of MRR showing self-coupling coefficient ($r$), cross-coupling coefficient ($k$) and round-trip transmission coefficient ($a$) where the ring is partially covered by a MoTe₂ flake. (b) The propagation loss ($\alpha_{TMDCs-Si}$) for TMDC-covered portion of the ring is found to be 0.4 dB/μm using cutback measurement. Tunability of (c) round-trip transmission coefficients explains the exponential decrease as a function of flake coverage. (d) Relationship between critical coverage (%) and power coupling co-efficient ($k_2$) assuming lossless coupling ($r_2+k_2=1$).

The change in resonant wavelength ($\lambda_{res}$) of a MRR is dependent on group index which takes into account the dispersion of the waveguide and can be defined as, $n_g = n_{eff} - \lambda_0 \frac{dn_{eff}}{d\lambda}$, here, for our case, since the resonance shift occurs over a narrow wavelength range, where $dn_{eff}/d\lambda \ll 1$, considering first order dispersion term, we approximate $n_g$ as $n_{eff}$ [30,31]. Therefore, the change in effective mode index ($\Delta n_{eff}$) is related to change in resonance ($\Delta \lambda$) following $\Delta n_{eff} = \frac{\Delta \lambda}{\lambda_{res}} * n_{eff,control}$, where, $n_{eff,control}$ is the effective mode index for Si MRR (i.e. without any TMDC flakes transferred). The effective index for the control sample (Si ring+SiO₂ cladding) can be found from FEM Eigenmode analysis choosing the TM-like mode in correspondence with our TM-grating designs used in measurements. We map out the resonance shift ($\Delta \lambda$) as a function of MoTe₂ coverage (Figure 4b, (i)) for needed calibration to determine the unknown index using a semi-empirical approach. The positive change in the effective index ($\Delta n_{eff}$) as a function of MoTe₂ coverage relates to an increased effective mode index with TMDCs transfer (Fig. 4b), (ii)) and corresponding red-shifts thereof.

Now, we obtain a resonance shift of 1 nm for 4 % coverage, which gives a corresponding change in effective index of 0.001 (Fig. 4b). At this point, it is important to establish the relation between the effective mode index with the refractive index of the unknown TMDCs. Since, in our case, the MRR is partially covered by MoTe₂ flakes and the resonance shift arising from change in effective mode index due to the change in coverage length, the effective refractive index of the ring can be formulated as an effective length-fraction index via,

$$n_{eff,ring} = \frac{(2\pi R - l)*n_{eff,control} + l*n_{eff}}{2\pi R} \quad (3)$$

where, $R$ is the radius of the ring and $l$ is the MoTe₂ coverage length. Using (3), it is straightforward to find $n_{eff}$ after TMDCs transfer.

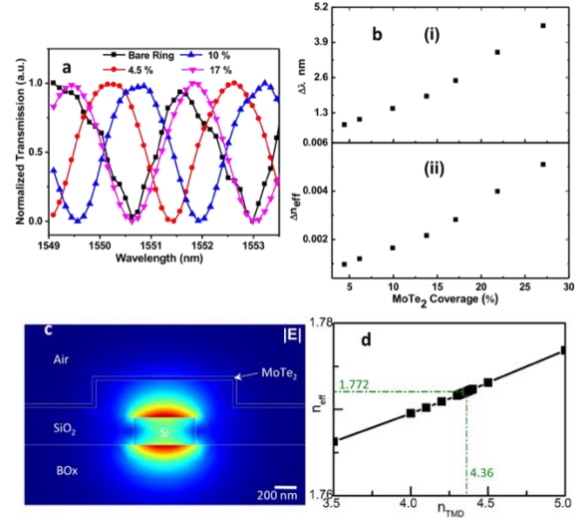

**Fig.4.** Ring resonator as refractive index sensor (a) Normalized transmission spectra for different coverages showing gradual red shift. Variation of (b) Resonance shift ($\Delta\lambda$) and effective index change ($\Delta n_{eff}$) extracted from (a) as a function of MoTe₂ coverage length. The mapped-out resonance shift as a function of coverage length provide the calibration curve to determine the unknown index of the TMDCs. (c) Mode profile (|E|) for the portion of the ring with MoTe₂ transferred flakes from Eigenmode analysis, and (d) FEM results for effective index, n_eff vs. the MoTe₂ index, n_TMDC to extract the material index from experimental results. The green dashed line exhibits the obtained MoTe₂ index from our results as 4.36.

Once, $n_{eff}$ is known, we can find the effective index of the heterogeneous optical mode (SOI plus TMDC) through Eigenmode analysis for a device geometry and provide experimentally measured thickness values of the flake (~35 nm). By sweeping the value of the TMDC ($n_{TMDCs}$ range = [3.5, 5]), material index in cross-sectional Eigenmode analysis of the waveguide structure in the MoTe₂ transferred section of the ring (Fig. 4c), we can match the numerically obtained effective index with that found from aforementioned experimental results. We find the index of the bulk MoTe₂ material to be 4.36 (Fig. 4d). This index value is closely aligned with reported values in previous studies for bulk MoTe₂[32]. We also find the imaginary part of the material index from the cutback method varying flake sizes to be ~0.011. This semi-empirical approach has several advantages over conventional ellipsometric technique to determine the refractive index of the unknown materials; for instance, in ellipsometry, large uniform flakes (few mm to cm) are needed since the beam spot size is generally large, which are still challenging to obtain with uniform properties across such scales. However, for the method presented here, small TMDCs flakes (~500 nm) can be measured. In fact, the limit of this technique is not bounded by the MRR waveguide width, since partial coverage of the top waveguide is also possible to measure, but with linear phase shift scaling with respect to area covered. Besides this, to determine refractive index of the materials from the ellipsometry data, the experimental data needs to be fitted by an exact physical model which bears ambiguity and thus introduces additional inaccuracy.

The technology outlook for heterogeneous integrating 2D materials with photonic integrated circuits is to develop next generation optoelectronic components such as modulators and photodetectors. These universal building blocks can be assembled in circuits for i) data links and interconnects including hybrid-technology options [33] including dual-function optoelectronic devices [34], ii) photonic integrated network-on-chip technology [35], iii) photonic digital-to-analog converters [36], and photonic analog processors such as iv) photonic residue arithmetic processors relying on cross-bar routers [37, 38], v) analog partial differential equation solvers [39], vi) photonic integrated neural networks [40-43], vii) photonic reprogrammable circuits such as content-addressable memories [44], or for optical phase-arrays in beam steering [45]. We note, that two dimensional materials, such as graphene can be grown in a foundry-near method such as using chemical vapor deposition (CVD) and subsequent etching. could be deployed for high-speed while preserving high modulation voltage efficiency showing <fJ/bit modulation efficiency [46] and the potential for 100 GHz fast modulation [47]. Scaling laws for opto-electronics and cavity options could be used further to reduce energy consumption of next generation 2D material based modulators [48, 49].

In conclusion, we have studied the interaction between few layers of $MoTe_2$ and Si MRR for the first time. We observed tunability of the coupling strength i.e. the ring resonator can be tuned from the over-coupled to the under-coupled regime while passing through the critically-coupled point. We further demonstrated a semi-empirical approach to determine the index of miniscule (~500 nm) TMDCs material flakes using the index sensitivity of the ring resonator. These findings along with the developed methodology for placing MRRs into critical coupling for active device functionality and determining the refractive index of 2D materials could be useful tools in future heterogeneous integrated photonic and optoelectronic devices. This developed technique could also be used to determine the optical refractive index of monolayer 2D materials, which is challenging with traditional techniques due to lateral TMDCs flake size, and atomic thickness of these materials.

**Funding.** National Science Foundation (NSF) (DMREF 14363300/1455050); AFOSR (FA9550-17-1-0377); ARO (W911NF-16-2-0194).